\documentstyle[preprint,aps,epsfig]{revtex}
\tightenlines
\begin{document} 
\draft
\title{\begin{flushright}
          {\small IFT-P. 008/99 \,\, gr-qc/9901006}
       \end{flushright}
       Scalar radiation emitted from a  source rotating 
       around a black hole}


\author{Lu\'\i s C.B. Crispino$^{1,2}$, Atsushi Higuchi$^3$, 
       and George E.A. Matsas$^1$}
\address{$^1$Instituto de F\'\i sica Te\'orica, 
         Universidade Estadual Paulista,\\
         Rua Pamplona 145, 01405-900, S\~ao Paulo, S\~ao Paulo, Brazil\\ 
         $^2$Departamento de F\'\i sica, Universidade Federal do Par\'a,\\ 
         Campus Universit\'ario do Guam\'a, 66075-900, Bel\'em, Par\'a,
         Brazil\\
         $^3$Department of Mathematics, University of York,\\ 
         Heslington, York YO10 5DD, United Kingdom}
\date{\today}
\maketitle 


\begin{abstract}
We analyze the scalar radiation emitted from a source rotating
around a Schwarzschild black hole using the framework of quantum field theory
at the tree level. We show that for relativistic circular
orbits the emitted power is about 20\% to 30\% smaller
than what would be obtained in Minkowski spacetime.
We also show that most of the emitted energy escapes to infinity.
Our formalism can readily be adapted to investigate similar processes.
\end{abstract}



\maketitle

\section{Introduction}
\label{sec:introduction}
Observational confirmation of the existence of black holes is
one of the most important challenges in astrophysics.
Recently a number of compact objects in X-ray binary systems
have been identified as black holes
since a careful analysis has shown that their masses are far
beyond any limit accepted for dead stars in general relativity~\cite{evid1}.
There also exists indirect evidence of the presence
of supermassive black holes in the centre of some galaxies~\cite{evid2}.
Nevertheless, unambiguous confirmation of the existence
of black holes would require the observation of effects due to the event
horizon itself. This is expected to be achieved  through precise
measurements of the electromagnetic radiation emitted from
black hole accretion disks (see e.g. \cite{Th,NY94}), and from the
gravitational radiation emitted from companion stars orbiting
black holes (see e.g. \cite{waves}).
Because radiation from sources orbiting black holes
plays such a crucial role in modern astrophysics and
because increasingly precise measurements
are leading to the observation of relativistic effects
occurring in the vicinity of the horizon~\cite{T}, 
investigation of how radiation-emission processes are modified
by the nontrivial curvature and topology of the black-hole
spacetime is particularly important.

In this paper we analyze analytically and numerically the 
{\em scalar} radiation emitted by
a source rotating around a black hole using the framework of
quantum field theory at the tree level, and compare the
results with those obtained in Newtonian gravity and in a theory
associated with one-graviton exchange in flat spacetime.
We show that these results coincide
asymptotically, as expected,
but considerably differ  close to the last stable circular orbit.
We also calculate the amount of emitted energy which is not absorbed
by the black hole.
The paper is organized as follows. In section~\ref{sec:GF} we present
the framework of quantum field theory in which we will work.
In section~\ref{sec:results} we obtain analytical and numerical
results for the emitted power.
In section~\ref{sec:CFT} we establish a connection between our
quantum field theory approach and classical field theory.
In section~\ref{sec:CxF} we compare our curved spacetime calculation
with flat spacetime calculations by considering (i) Newtonian
gravity and (ii) the one-graviton exchange theory. 
In Section~\ref{sec:asymptotics} we calculate analytically and
numerically the amount of radiation which is not absorbed by the hole.
Finally in  section~\ref{sec:final} we make some remarks on our results.  
We use natural units
$c = \hbar = G = 1$ and signature $(+---)$ throughout this paper.

\section{General framework}
\label{sec:GF}
A nonrotating black hole with mass $M$ is described
by the Schwarzschild line element
\begin{equation}
ds^2 = f(r) dt^2 - f(r)^{-1} dr^2
       - r^2 d\theta ^2 - r^2 \sin^2{\theta}\, d\phi^2 \;\; ,
\label{LiEl}
\end{equation}
where $f(r) = 1-2M/r$.
Let us consider a circularly moving scalar source
at $r = R_S$ with constant angular velocity $\Omega>0$ (as measured by
asymptotic  static observers) on the plane $\theta = \pi /2$
described by
\begin{equation}
j(x^\mu) =
\frac{q}{{\sqrt {-g}} \; u^0}
\delta (r-R_S) \; \delta (\theta - \pi /2) \; \delta (\phi - \Omega t)
\label{j}
\end{equation}
with $g \equiv det(g_{\mu \nu} )$, where the constant $q$ determines
the magnitude of the source-field coupling. The four-velocity of this
source is
\begin{equation}
u^\mu (\Omega , R_S)=
[(f(R_S)-R_S^2 \Omega^2)^{-1/2},0,0, \Omega/(f(R_S)-R_S^2\Omega^2)^{1/2}]\;\;.
\end{equation}
We have normalized the source $j$ by requiring
that $\int d\sigma  j(x^{\mu}) = q $,
where $d\sigma $ is the proper three-volume element
orthogonal to $u^\mu$.
Let us now minimally couple $j(x^{\mu})$ to a massless scalar field
$\hat \Phi (x^{\mu})$ so that the total Lagrangian density is
\begin{equation}
{\cal L} = \sqrt{-g} \, \left( {1 \over  2}
\nabla^{\mu} \hat \Phi \nabla_{\mu} \hat \Phi
 +j \hat \Phi \right) \; .
\label{Lag}
\end{equation}

\begin{figure}
\begin{center}
\rm{\epsfig{file=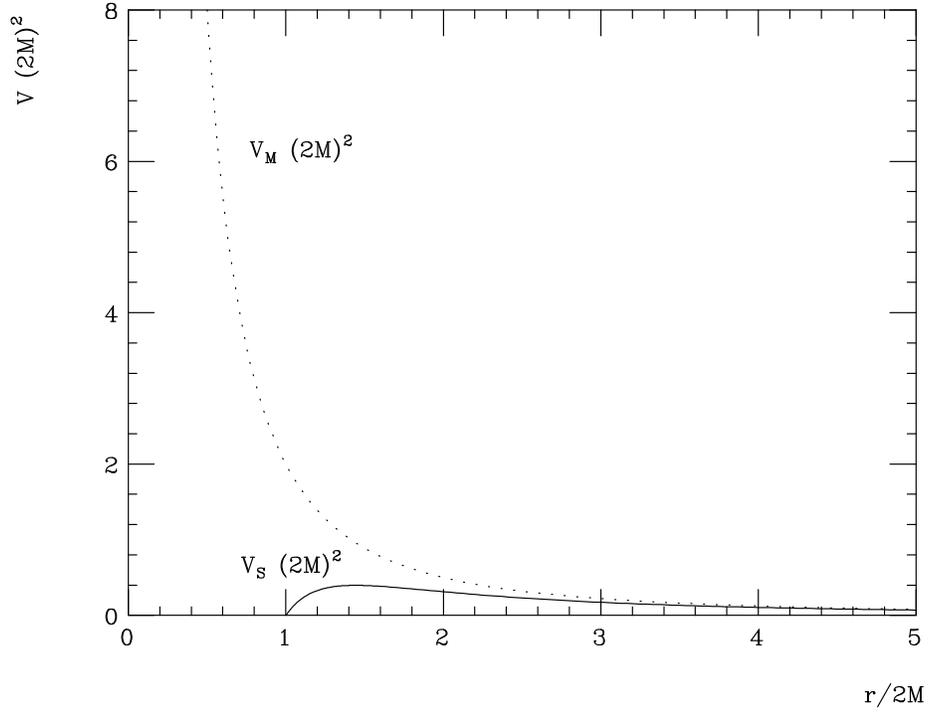,width=0.7\textwidth,angle=90}}
\end{center}
\vskip -1 cm
\caption{ Scattering potentials $V_M$ and $V_S$ are
plotted as functions of $r/2M$ for $l=1$,
where we recall that $V_M$ and $V_S$ are defined as
functions of Minkowski and Schwarzschild
$r$ coordinates, respectively.
Asymptotically $V_M$ and $V_S$
fall as $1/r^2$. $V_S$ is only defined outside
the black hole ($r>2M$). Because of the nonexistence of the
event horizon in Minkowski spacetime, $V_M$ is also defined
in the region $0<r\le 2M$.}
\label{fig01}
\end{figure}

The positive-frequency modes in spherically symmetric static spacetime can
be given in the form
\begin{equation}
u_{\omega l m}(x^\mu)
= \sqrt{\frac{\omega}{\pi}}\,
\frac{\psi_{\omega l} (r)}{r} Y_{l m} (\theta, \phi) e^{-i\omega t}
\;\;\; (\omega>0) \;\; ,
\label{Udef}
\end{equation}
where $\Box u_{\omega l m} = 0$. In the present case
the functions $\psi_{\omega l} (r) $ satisfy the differential equation
\begin{equation}
\left[ -f(r)
\frac{d}{dr}\left(f(r)
\frac{d\ }{dr} \right) + V_{S}(r) \right]
\psi^S_{\omega l}(r)
= \omega^2 \psi^S_{\omega l}(r)
\label{RPWES}
\end{equation}
with the following scattering potential (see figure~\ref{fig01}):
\begin{equation}
V_{S}(r) = \left( 1-2M/r\right)
\left[ 2M/r^3 + l(l+1)/r^2\right] \;\; .
\label{VSc}
\end{equation}
In terms of the
dimensionless tortoise coordinate $x \equiv r/2M +\ln (r/2M-1)$,
(\ref{RPWES}) can be rewritten as
\begin{equation}
\left[ -{{d^2} \over {d x^2}} +
 4M^2 V_S [r(x)]\right]\psi^S_{\omega l}
= 4M^2\omega^2\psi^S_{\omega l} \;\; .
\label{KG2}
\end{equation}

The scalar field  can be expanded in terms of the complete set
of positive- and negative-frequency modes  as
\begin{equation}
\hat \Phi^{\rm in}(x^\mu) =
\sum_{l=0}^{\infty} \sum_{m=-l}^{l} \int_0^{\infty} d\omega
\left[ u_{\omega l m}(x^\mu) a^{\rm in}_{\omega l m}
+ H.c.
\right] \;\; .
\label{field}
\end{equation}
The Klein-Gordon inner product $\sigma_{KG}$ is defined by
\begin{equation}
\sigma_{KG}(\phi,\psi) = i \int_{\Sigma_t}
d\Sigma\, n^\mu (\phi^*\nabla_{\mu}\psi - \nabla_\mu\phi^*\cdot
\psi)\;\;,  \label{KlGd}
\end{equation}
where $n^{\mu}$ is the future-pointing unit vector orthogonal to the
Cauchy surface $\Sigma_t$ with $t = const$.
The modes $u_{\omega l m}$
are Klein-Gordon orthonormalized:
\begin{eqnarray}
\sigma_{KG}(u_{\omega l m},u_{\omega'l'm'}) & = & \delta(\omega - \omega')
\delta_{ll'}\delta_{mm'}\;\;,
\label{KG} \\
\sigma_{KG}(u_{\omega l m}^*,u_{\omega'l'm'}) & = & 0\;\;.
\end{eqnarray}
Then the creation and annihilation operators,
$a^{{\rm in}\dagger}_{\omega l m}$ and
$a^{\rm in}_{\omega l m}$,
satisfy the usual commutation relations
$
[ a^{\rm in}_{\omega l m},
a^{{\rm in}\dagger}_{\omega' l' m' }]
=
\delta(\omega - \omega')
\delta_{l l'}
\delta_{m m'} .
$

We see from (\ref{KG2}) that close to and far away from the horizon,
the modes $\stackrel{\rightarrow}{\psi^S_{ \omega l}} (x)$
purely incoming from the past horizon $H^-$, and the modes
$\stackrel{\leftarrow}{\psi^S_{ \omega l}} (x)$
purely incoming from the past null infinity ${\cal J^-}$
can be written as~\cite{HMSPRD97,HMSPRD98}
\begin{equation}
\stackrel{\rightarrow}{\psi^S_{ \omega l}} (x) \approx
\left\{
\begin{array}{l}
A_{ \omega l} (e^{2 i M\omega x} +
\stackrel{\rightarrow}{{\cal R}}_{ \omega l}
e^{- 2i M\omega x})
\ \ \ ( r{\raise0.3ex\hbox{$\;>$\kern-0.75em\raise-1.1ex\hbox{$\sim\;$}}}2M)
\;\; ,
\label{psiIa}
\\
2 i^{l+1} A_{ \omega l} \stackrel{\rightarrow}{{\cal T}}_{ \omega l}
M\omega x \; h_l^{(1)} ( 2M\omega x)
\ \ \  ( r \gg 2M) \;\; ,
\label{->}
\end{array}
\right.
\end{equation}
and
\begin{equation}
\stackrel{\leftarrow}{\psi^S_{ \omega l}} (x) \approx
\left\{
\begin{array}{l}
{A}_{ \omega l} \stackrel{\leftarrow}{{\cal T}}_{ \omega l} e^{-2i M\omega x}
\ \ \ ( r{\raise0.3ex\hbox{$\;>$\kern-0.75em\raise-1.1ex\hbox{$\sim\;$}}}2M)
\;\; ,
\label{psiCPa} \\
2 A_{ \omega l}  M\omega x  [ (-i)^{l+1} h_l^{(2)} ( 2M\omega x)
+ {i^{l+1}}  \stackrel{\leftarrow}{{\cal R}}_{ \omega l} h_l^{(1)}
( 2M\omega x) ]
\ \ \  ( r \gg 2M) \;\; ,
\label{<-}
\end{array}
\right.
\end{equation}
where $h_l^{(i)}$ $(i=1,2)$ are spherical Bessel functions
of the third kind~\cite{AS} and the overall normalization constant
is determined by (\ref{KG}) as
$A_{ \omega l}  =  (2\omega)^{-1}$.

The emitted power is given by
\begin{equation}
W^{em}_{l m} =
\int^{+ \infty}_0 d\omega \, \omega \;
{\left| {\cal A}^{em}_{\omega l m} \right| ^2}/{T} \;\; ,
\label{Wem}
\end{equation}
where
\begin{equation}
        {\cal A}^{em}_{\omega  l m } = \;  \langle \omega  l m |
        \; i \int d^4 x {\sqrt {-g}} \; j (x^\mu) \hat\Phi (x^\mu) \;
        | 0 \rangle \;
        = i\int d^4 x \sqrt{-g}\,j(x^\mu)u_{\omega lm}^*(x^\mu)
\label{Amplitude}
\end{equation}
is the emission amplitude at the tree level, and $T=2\pi \delta(0)$
is the total time  measured by asymptotic observers~\cite{IZ}.
We have chosen the initial state $| 0 \rangle$ to be the Boulware vacuum,
i.e. $a^{\rm in}_{\omega l m}| 0 \rangle=0$.
If we had chosen the Unruh or Hartle-Hawking vacuum~\cite{I},
then (\ref{Wem}) would be associated with the {\em net}
radiation emitted from the source since the absorption and stimulated
emission rates (which are induced  by the presence of thermal fluxes)
are exactly the same. Note that
for sources in constant circular motion the amplitude
${\cal A}^{em}_{\omega  l m }$ is proportional to
$\delta (\omega - m \Omega)$.
Hence the frequency  of emitted waves is constrained by
$\omega =m \Omega$. In particular, since $\Omega >0$, no
waves with $m \leq 0$
are emitted.

\section{Emitted power: numerical and analytical results}
\label{sec:results}
The general solution of (\ref{KG2}) is not
easy to analyze~\cite{JC} and therefore we first estimate
the radiated power numerically. (Later we will compare it 
with the one obtained with our analytic approach.)
To do so we solve (\ref{KG2}) for the left- and right-moving
radial functions $\stackrel{\leftarrow}{\psi^S_{ \omega  l}} (r)$ and
$\stackrel{\rightarrow}{\psi^S_{ \omega  l}} (r)$
with asymptotic boundary conditions
compatible with equations (\ref{->}) and  (\ref{<-}).
The method for finding the 
the functions $\stackrel{\leftarrow}{\psi^S_{ \omega l}}$ can be summarized
as follows.   
(The functions $\stackrel{\rightarrow}{\psi^S_{ \omega  l}}$ are
obtained in a similar manner.)
We recall from (\ref{<-}) that close to the horizon
$\stackrel{\leftarrow}{\psi^S_{ \omega  l}}
\propto \exp (-2iM\omega x)$. Thus we construct the solution 
${\chi_{ \omega  l}}(x)$ of (\ref{KG2}) with fixed $\omega$ satisfying 
${\chi_{ \omega  l}}(x) = \exp (-2iM\omega x) $
as the initial condition for $x <0, |x| \gg 1$ by evolving it
numerically towards large $x$. 
Far away from the horizon we have 
${\chi_{ \omega  l}}(x)
\approx A \exp (-2iM\omega x)+ B \exp (2iM\omega x)$ [see (\ref{<-})
and recall that 
$h_l^{(1)} (2M\omega x) 
\approx (-i)^{l+1} e^{2iM\omega x}/ (2M\omega x)$
and
$h_l^{(2)} (2M\omega x) 
\approx i^{l+1} e^{-2iM\omega x}/ (2M\omega x)$ for $2M\omega x \gg 1$].
The constants $A$ and $B$ (with $|A|^2 - |B|^2 =1$) 
are determined from the numerically obtained values of
$ {\chi_{ \omega l }}(x)$ and 
$ {d\chi_{ \omega l }}/dx$ for $2M\omega x \gg 1$.
Now, by requiring the asymptotic boundary conditions
compatible with (\ref{<-}), we find 
$
\stackrel{\leftarrow}{\psi^S_{ \omega  l}} (x)
=
{\chi_{ \omega  l}} (x)/(2\omega A) .
$
(In particular,
$\stackrel{\leftarrow}{{\cal T}}_{ \omega l} = 1/A$
and
$\stackrel{\leftarrow}{{\cal R}}_{ \omega l} = B/A$.)

\begin{figure}
\begin{center}
\rm{\epsfig{file=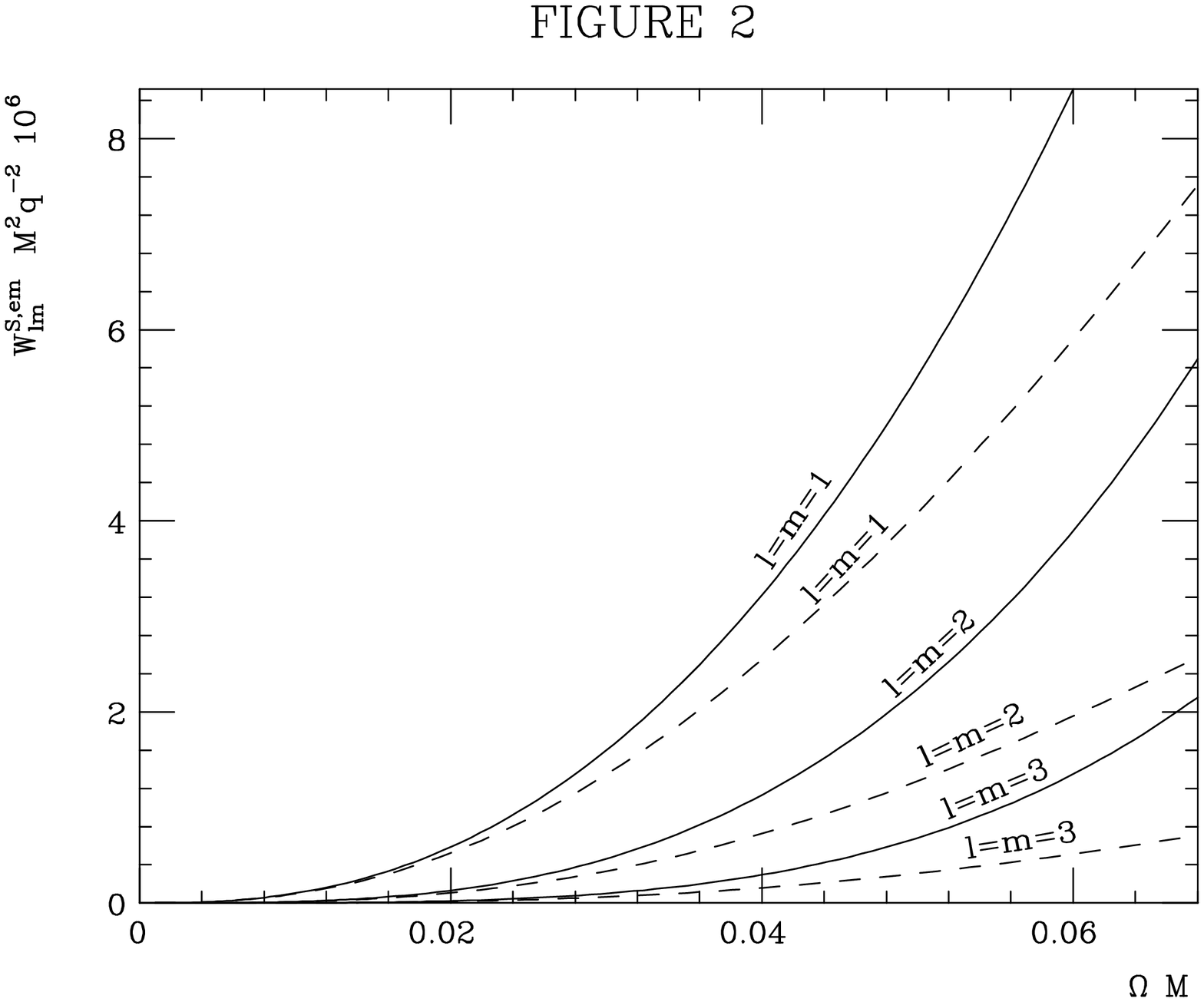,width=0.7\textwidth,angle=90}}
\end{center}
\vskip -1 cm
\caption{ The radiated power $W^{S,em}_{lm}$ is plotted as a function of
$\Omega$ for geodesic orbits. Solid and dashed lines are
associated with the
numerical calculation and analytic approximation, respectively.
The maximum $ \Omega M $
considered is $0.068$ which is associated with the last stable
circular orbit. As expected our analytic approximation
is only accurate to describe the  emission of low-energy particles
as it can be seen
from the coincidence of the numerical and (approximate) analytic curves
for small $\Omega$.
It is clear that small angular-momentum waves
give the main contribution to the total radiated power.}
\label{fig02}
\end{figure}

By substituting the radial functions 
$ \stackrel{\leftarrow}{\psi^S_{ \omega  l}}$ and 
$ \stackrel{\rightarrow}{\psi^S_{ \omega  l}}$ in (\ref{Udef}) 
to construct the
right- and left-moving modes $\stackrel{\rightarrow}{u}_{\omega lm}$ and
$\stackrel{\leftarrow}{u}_{\omega lm}$, respectively, and using
(\ref{Wem}) with (\ref{Amplitude}), we find the corresponding
radiated powers:
\begin{equation}
\stackrel{\rightarrow}{W}_{lm}^{S,em}
= 2m^2\Omega^2 q^2[f(R_S) - R_S^2\Omega^2] \;
| \stackrel{\rightarrow}{\psi^S_{\omega_0 l}}(R_S)/R_S |^2 \;
|Y_{lm}(\pi/2,\Omega t)|^2\;\;,
\label{Wnum}
\end{equation}
where $\omega_0 \equiv m\Omega$ and $l,m \geq 1$, and similarly for
$\stackrel{\leftarrow}{W}_{lm}^{S,em}$.
We note~\cite{GR} that $Y_{lm}(\pi/2,\Omega t) = 0$ if $l+m$ is odd and
\begin{equation}
        |Y_{l m}(\pi/2,\Omega t)|^2 =
\frac{2l+1}{4\pi}\frac{(l+m-1)!!(l-m-1)!!}{(l+m)!!(l-m)!!}
\label{Y}
\end{equation}
if $l+m$ is even, which is of course time independent.  We have defined
$n!! \equiv n(n-2)\cdots 1$ if $n$ is odd and $n!! \equiv n(n-2)\cdots 2$
if $n$ is even and $(-1)!! \equiv 1$.
Now, the condition that our source be in a stable circular
geodesic implies
$R_S = (M\Omega^{-2})^{1/3}$ as is well known~\cite{W}. By using this formula,
$\stackrel{\rightarrow}{W}_{lm}^{em}$ and
$\stackrel{\leftarrow}{W}_{lm}^{em}$ can be cast
as functions of quantities measured at infinity alone,
namely, $\Omega$ and $M$.
In figure~\ref{fig02} we plot the total radiated power with fixed angular
momentum,
$
W^{S,em}_{l m} = \; \stackrel{\rightarrow}{W}^{S,em}_{l m}
+ \stackrel{\leftarrow}{W}^{S,em}_{l m}\;
$,
as a function of the angular velocity $\Omega$ for different values
of $l$ and $m$. Note that
${W}^{S,em}_{l m}=0$ for odd $l+m$ because $Y_{lm}(\pi/2,\Omega t) = 0$ in
this case.

Next we consider an analytic
approximation valid for low-frequency modes.  Let us
first recall that waves emitted from circularly moving sources
obey the constraint $\omega=m \Omega$.
Thus, waves with $m = 1$ (which turn out to be the
most important ones for our purposes) emitted from sources in stable
circular geodesic orbits ($R_S>6M$, $\Omega = \sqrt{M/ R_S^3} $\,)
have maximum frequency
$\omega^{\rm max} = (6\sqrt{6}\; M)^{-1}\ll \sqrt{V_S^{\rm max}} $,
where $V_S^{\rm max}$ is the maximum of the scattering potential $V_S$.
Hence waves with small angular momentum have
small frequencies in comparison with $\sqrt{V_S^{\rm max}}$, i.e.,
$\omega^2/ V^{\rm max}_S < {(\omega^{\rm max})}^2/ V^{\rm max}_S \approx
4 \times 10^{-2} \ll 1$ (see figure~\ref{fig01}).
A similar analysis for waves with arbitrary $m$ shows that
$\omega^2/ V^{\rm max}_S < 10^{-1}$.
Therefore, we will consider the approximation where the radial
functions are replaced by  their leading terms for small $\omega$
(see equations (6.6) and (7.2) in \cite{HMSPRD98}, and
\cite{misprint} for a correction):
\begin{equation}
\stackrel{\rightarrow}{\psi^S_{\omega l}} (r)
\approx 2 r Q_l(r/M-1)
\label{Q}
\end{equation}
and
\begin{equation}
\stackrel{\leftarrow}{\psi^S_{\omega  l}} (r)
\approx \frac{2^{2l} (l!)^3 (M \omega)^l r P_l(r/M-1)}{(2l)!
\, (2l+1)!} \;\; ,
\label{P}
\end{equation}
where $P_l(x)$ and $Q_l(x)$ are the Legendre functions.
By substituting $\stackrel{\rightarrow}{\psi^S_{ \omega l}}$
and  $\stackrel{\leftarrow}{\psi^S_{ \omega l}}$
(given in equations (\ref{Q}) and (\ref{P}) respectively)
in (\ref{Wnum}) and a similar expression for
$\stackrel{\leftarrow}{W}_{lm}^{em}$, we find
the corresponding radiated powers [$R_S = (M\Omega^{-2})^{1/3} $]:
\begin{equation}
\stackrel{\rightarrow}{W}^{S,em}_{l m} \approx
8 q^2 m^2 \Omega^2
\, (f(R_S) - R_S^2 \Omega ^2) \,| Q_l(R_S/M-1)|^2
\; |Y_{lm}(\pi/2,\Omega t)|^2
\label{WlmSU}
\end{equation}
and
\begin{equation}
\stackrel{\leftarrow}{W}^{S,em}_{l m} \approx
 \frac{2^{4l+1} q^2 (l!)^6 m^{2l+2} M^{2l} \Omega^{2l+2}}
{[(2l)!]^2 [(2l+1)!]^2}
\, (f(R_S) - R_S^2 \Omega ^2) \, | P_l(R_S/M-1) |^2
\; |Y_{lm}(\pi/2,\Omega t)|^2 \;\;
\label{WlmSCP}
\end{equation}
where $l,m \geq 1$.
The total radiated power with fixed angular momentum
obtained in this approximation
is also shown in figure~\ref{fig02}.
We see from this figure that our analytic approximation
has better accuracy for small $\Omega$ as expected. [Note that 
$\omega \propto \Omega$ and see discussion above (\ref{Q})].

It can be  seen from figure~\ref{fig02} that the total radiated power
\begin{equation}
W^{S,em} = \sum_{l=1}^{\infty} \sum_{m=1}^{l} {W}^{S,em}_{l m}
\label{WHH}
\end{equation}
will be dominated by waves with small $l$.
For circular geodesic orbits far enough from
the horizon  we can use equations (\ref{WlmSU})-(\ref{WHH})
to write the radiated power in a simple form:
\begin{equation}
\left. W^{S,em} \right|_{R \gg r_S} \approx
{q^2 M^{2/3} \Omega^{8/3}}/{12 \pi} \;\;  .
\label{WSHH}
\end{equation}

\section{Connection with classical field theory}
\label{sec:CFT}
Since our calculations are performed
at the tree level, our results can be interpreted in classical
field theory as follows (see, e.g., \cite{HM93}).
One can show that the energy of a classical field $\phi$ can be
written as
\begin{equation}
E = \frac{i}{2}\sigma_{KG}(\phi,\partial_t\phi)\;\;, \label{Energy}
\end{equation}
where the Klein-Gordon inner product $\sigma_{KG}$ is defined by
(\ref{KlGd}). 
The classical field generated by a
source $j(x^\mu)$ can be expressed in general as
\begin{equation}
\phi(x^\mu) =
i \sum_{l=0}^\infty\sum_{m=-l}^{l}\int_0^{+\infty}
d\omega \left[ {\cal A}_{\omega lm}^{em} u_{\omega lm}(x^\mu)
- {\cal A}_{\omega lm}^{em*}u_{\omega lm}(x^\mu)^* \right]\;\;,
\end{equation}
where ${\cal A}_{\omega lm}^{em}$ is defined by equation (\ref{Amplitude}).
By substituting this in (\ref{Energy}) and using the orthonormality of
$u_{\omega lm}$ we obtain the energy as
\begin{equation}
E = \sum_{l=0}^\infty\sum_{m=-l}^{l}\int_0^{+\infty}d\omega\;
\omega |{\cal A}_{\omega lm}^{em}|^2\;\;,
\end{equation}
which agrees with (\ref{Wem}).
(See, e.g., \cite{Pois} for other classical-field-theory
analyses mainly developed to study gravitational wave emission.)

\section{Curved vs. flat spacetime calculations}
\label{sec:CxF}
Now we compare the radiated power  $W^{S,em}$ in Schwarzschild
spacetime with that
in Minkowski spacetime.
We consider radiation from a source in
circular motion in Minkowski spacetime due to the presence of
some gravitational force.  The latter should
give  fairly good results for the case of a source rotating around a
star that is not very dense, but not for the case
of a source rotating close to
a black hole as will be shown.  

Let us represent the scalar source
in Minkowski spacetime by
\begin{equation}
j^M (x^\mu) =
\frac{q}{R_M^2  \; \gamma }
\delta (r-R_M)\; \delta (\theta - \pi /2)\; \delta (\phi - \Omega t) \;\; ,
\label{jM}
\end{equation}
where $\gamma = 1/\sqrt{1 - R_M^2 \Omega^2} $.
We are working here with spherical coordinates defined
through the Minkowski line element given by
(\ref{LiEl}) with $f(r)=1$.
Note that the normalization of this source is chosen so that
$\int d\sigma\,j^M(x^\mu) = q$, where $d\sigma$ is the proper
three-volume element orthogonal to the world line of the source, as in
the Schwarzschild case.

It is not {\it a priori}
meaningful to compare the radiated powers from the sources
in Schwarzschild and Minkowski spacetimes with the same value of $r$.
For this reason,
our results will be expressed in terms of $\Omega$. It is meaningful
to compare the radiated powers using $\Omega$ because this is a
(coordinate free) quantity measured by asymptotic observers.
We expand the scalar field as in equation (\ref{field}) with Klein-Gordon
orthonormalized
positive-frequency modes given by
(\ref{Udef}), where the radial functions
are given by
$
\; \psi^M_{\omega l} (r) = r j_l (\omega r)
$
(see, e.g., \cite{HMSPRD98}),
which satisfy  the differential equation
\begin{equation}
\left( - d^2/dr^2  + V_{M} \right) \psi^M_{\omega l}(r)
=  \omega^2 \psi^M_{\omega l}(r) \;\;
\label{RPWEMV}
\end{equation}
with
$
\; V_{M} \equiv {l (l+1)}/{r^2} \; .
$
The total radiated power calculated in Minkowski
spacetime is
\begin{equation}
W^{M,em} = \sum_{l=1}^{\infty} \sum_{m=1}^{l}
           2 q^2 m^2 \Omega^2 \gamma ^{-2}  \,
           | j_l (m \Omega R_M)|^2   |Y_{lm}(\pi/2,\Omega t)|^2 \;\; ,
\label{WemM}
\end{equation}
where we have used the Minkowski vacuum as the initial state.
In order to compare this expression with the
radiated power obtained through classical field theory, we
adapt the standard derivation of Larmor's formula for electric
charges (see, e.g., \cite{J}) to the case of scalar sources:
\begin{equation}
W^{M, em}_{class} = {q^2 a^2}/{12 \pi}  \;\;  ,
\label{WL}
\end{equation}
where $a$ is the  proper acceleration of the source
with an arbitrary trajectory.
For circular orbits, we have $a  = \gamma ^2 \Omega ^2 R_M $.
The equality $W^{M,em} = W^{M,em}_{class}$ follows from the
following formula: 
\begin{equation}
\sum_{l=1}^\infty\sum_{m=1}^l m^2[j_l(mz)]^2|Y_{lm}(\pi/2,\phi)|^2
= \frac{1}{24\pi}\frac{z^2}{(1-z^2)^3}\;\;, \label{summation}
\end{equation}
for $|z| < 1$.  A proof of this formula is given in the Appendix.  We also
verified it numerically for a wide range of values of $z$. 

\begin{figure}
\begin{center}
\rm{\epsfig{file=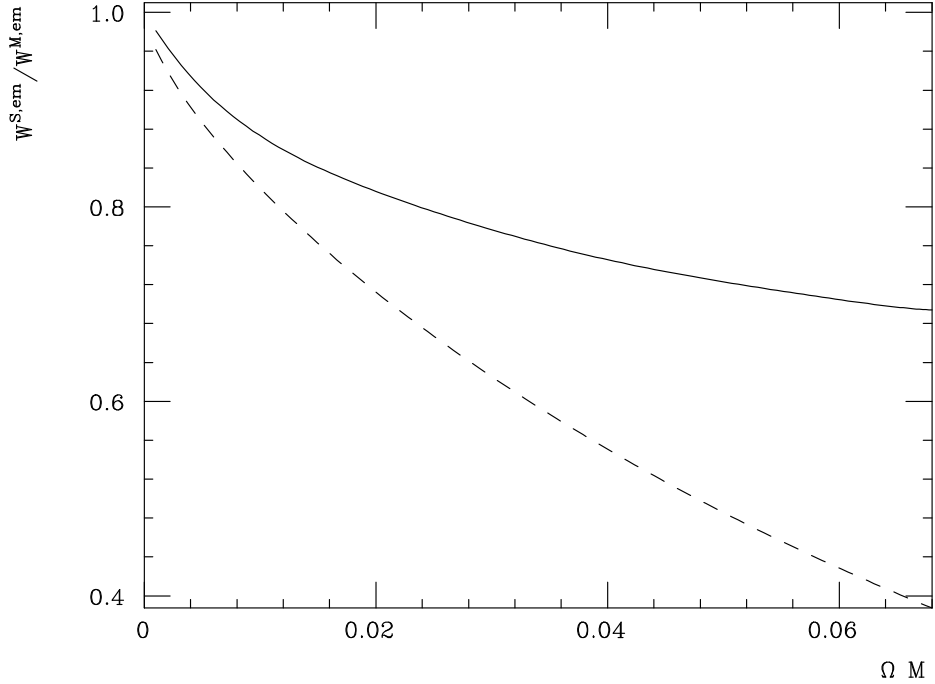,width=0.7\textwidth,angle=90}}
\end{center}
\vskip -1 cm
\caption{ The $W^{S,em}/W^{M,em}$ ratio
is plotted as a function of $ \Omega$
as measured by asymptotic observers,
where the summations involved in the calculation of $W^{S,em}$ and
$W^{M,em}$ were performed up to $l=3$.
$W^{S,em}$ and $W^{M,em}$ are the power emitted by an
orbiting source as calculated by asymptotic observers who assume
Schwarzschild spacetime and Minkowski spacetime with
Newtonian gravity respectively. The graph is plotted up to
$\Omega M = 0.068$, since this is the faster stable circular
orbit according to General Relativity.
Solid and dashed lines are
associated with the numerical calculation and analytic approximation,
respectively.
Asymptotically ($\Omega \to 0$)
we have $W^{S,em}/W^{M,em} \to 1$. As $\Omega$ increases,
however, we see that $W^{S,em}/W^{M,em}$ decreases.}
\label{fig03}
\end{figure}

\begin{figure}
\begin{center}
\rm{\epsfig{file=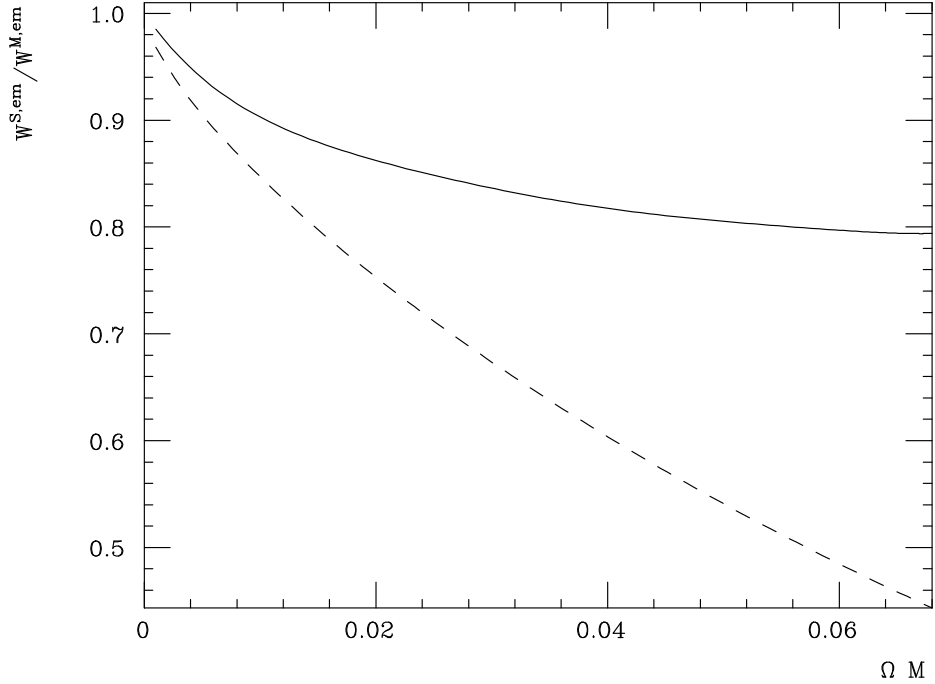,width=0.7\textwidth,angle=90}}
\end{center}
\vskip -1 cm
\caption{Similarly to figure~\ref{fig03}, the $W^{S,em}/W^{M,em}$
ratio is plotted as a function of $ \Omega$.
Here, however, $W^{M,em}$ is the power emitted by an
orbiting source as calculated by asymptotic observers who assume
Minkowski spacetime with
a gravitational force induced by gravitons, rather than with
the usual Newtonian force.
We note that figure~\ref{fig03} and figure~\ref{fig04} have
similar features:
(i) $W^{S,em}/W^{M,em} \to 1$ asymptotically ($\Omega \to 0$)
and,
(ii) $W^{S,em}/W^{M,em}$ decreases up to from 20\%
(see figure~\ref{fig04}) to 30\% (see figure~\ref{fig03})
as $\Omega$ increases.}
\label{fig04}
\end{figure}

Now, in order to compare $ W^{M,em}$ with $W^{S,em} $,
we shall cast $R_M$ as a function of $\Omega$ by imposing
the condition that the scalar source be in circular orbit due to the 
influence of a gravitational force. There is no unique way to define a
gravitational field in flat spacetime (see, e.g., \cite{MTW}).
Assuming Newtonian gravity and using
Kepler's third law
$R_M(\Omega) = (M \Omega^{-2})^{1/3}$
in (\ref{WemM}), we find $W^{M,em}$ as a function of $\Omega$.
We plot the ratio $W^{S,em}/W^{M,em}$ in figure~\ref{fig03}
as a function of $\Omega$.
A similar result is obtained if one considers the gravitational
force generated by the one-graviton exchange diagram~\cite{MTW}.
In this case a straightforward calculation gives
$\gamma^{2/3} R_M = (M \Omega^{-2})^{1/3}$ where $\gamma$ is
defined below equation (\ref{jM}). By solving this equation for $R_M$,
we obtain
\begin{eqnarray}
  R_M(\Omega)  & = &{{{2^{{1\over 3}}}\,{M^2}\,{\Omega^2}}\over
          {3\,{{\left( 27\,M\,{\Omega^4} - 2\,{M^3}\,{\Omega^6} +
          {\sqrt{-4\,{M^6}\,{\Omega^{12}} +
          {{\left( 27\,M\,{\Omega^4} - 2\,{M^3}\,{\Omega^6} \right) }^2}}}
          \right) }^{{1\over 3}}}}} +
\nonumber 
\\
      & + & {{{{\left( 27\,M\,{\Omega^4} - 2\,{M^3}\,{\Omega^6} +
          {\sqrt{-4\,{M^6}\,{\Omega^{12}} +
          {{\left( 27\,M\,{\Omega^4} - 2\,{M^3}\,{\Omega^6} \right) }^2}}}
          \right) }^{{1\over 3}}}}\over {{2^{{1\over 3}}}\;3\;{\Omega^2}}}
           - {{M}\over 3} \; ,
\label{RMG}
\end{eqnarray}
where $R_M > 0$.
We substitute
(\ref{RMG}) in (\ref{WemM}) to find the radiated power
$W^{M,em}$. We plot in figure~\ref{fig04}
the ratio $W^{S,em}/W^{M,em}$ as a function of $\Omega$.
Asymptotically, the radiated power is
\begin{equation}
\left. W^{M,em} \right|_{R \gg 2M} \approx
{q^2 M^{2/3} \Omega^{8/3}}/{12 \pi}  \;\;
\label{WTQC1}
\end{equation}
in either case.
Figure~\ref{fig03} and figure~\ref{fig04} show that
for $\Omega \to 0$ the ratio $W^{S,em}/W^{M,em}$
tends to unity as it should
[see equations (\ref{WSHH}) and (\ref{WTQC1})]. As the source approaches the
last stable orbit, however, our numerical and approximate analytic results
show that $W^{S,em}/W^{M,em} < 1$.  This is not a trivial
consequence of redshift
since the frequency $\Omega$ is measured at infinity in both the Schwarzschild
and Minkowski calculations.

\section{Asymptotic radiation}
\label{sec:asymptotics}
\begin{figure}
\begin{center}
\rm{\epsfig{file=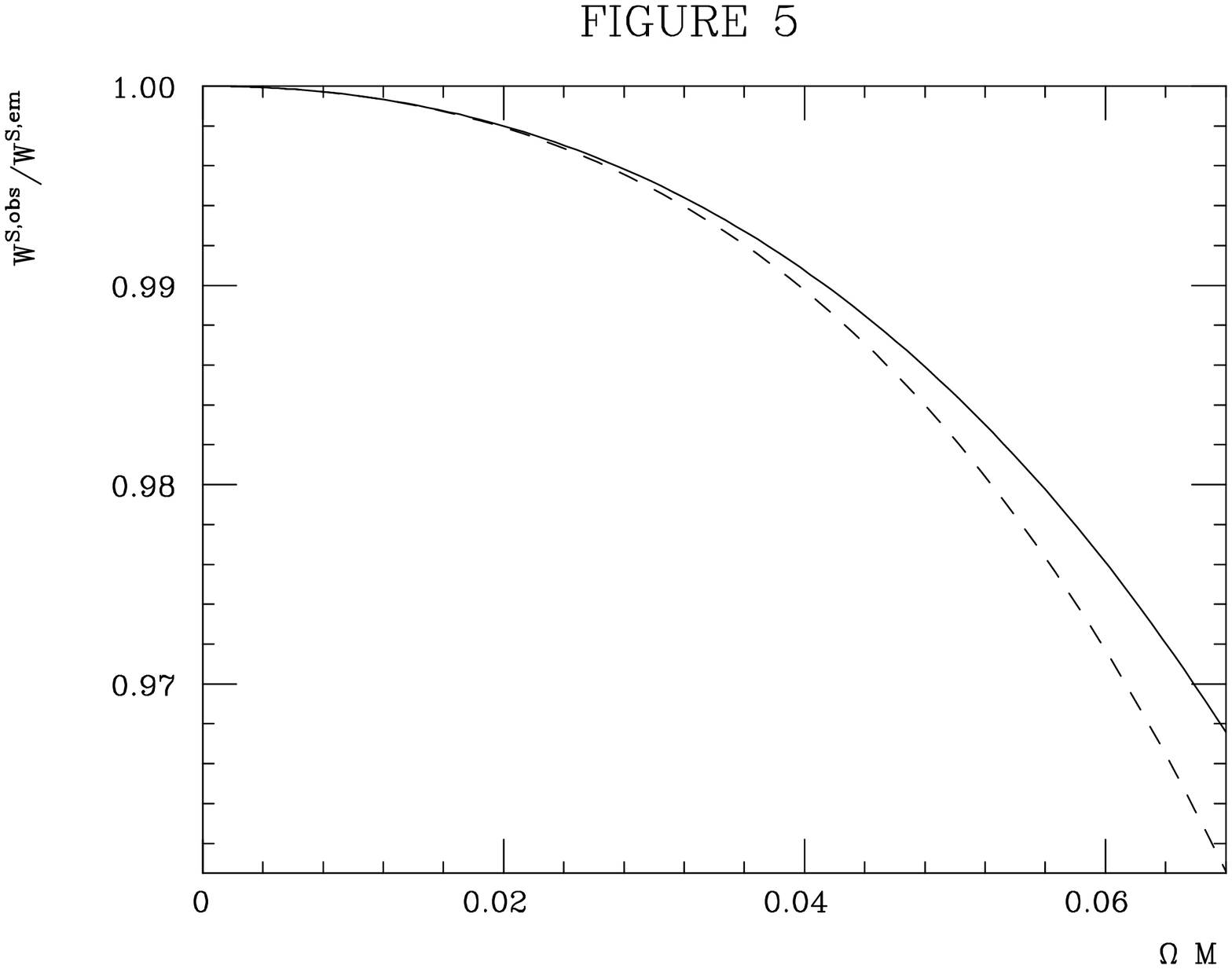,width=0.7\textwidth,angle=90}}
\end{center}
\vskip -1 cm
\caption{ The ratio $W^{S,obs}/W^{S,em}$
is plotted as a function of $ \Omega$ for geodesic orbits,
where the summations involved in the calculation of
$W^{S,obs}$ and $W^{S,em}$  were performed up to $l=3$.
Solid and dashed lines are
associated with the numerical calculation and analytic approximation,
respectively.
We note that most of the emitted energy
is radiated away to infinity.}
\label{fig05}
\end{figure}

Finally, it is interesting to compute
the amount of the radiated power that escapes to infinity, namely
\begin{equation}
W^{S,obs} = \sum_{l=1}^{\infty} \sum_{m=1}^{l}
            \left[|\stackrel{\rightarrow}{{\cal T}}_{ \omega_0 l}|^2
                   \stackrel{\rightarrow}{W}^{S,em}_{lm}  +
                  |\stackrel{\leftarrow}{{\cal R}}_{ \omega_0 l}|^2
                   \stackrel{\leftarrow}{W}^{S,em}_{lm} \right]  \;\; .
\label{WSO}
\end{equation}
By using
$
|\stackrel{\leftarrow}{{\cal R}}_{ \omega_0 l}|^2=
|\stackrel{\rightarrow}{{\cal R}}_{ \omega_0 l}|^2=
1-|\stackrel{\rightarrow}{{\cal T}}_{ \omega_0 l}|^2
$, the observed power (\ref{WSO}) can be
written as
\begin{equation}
W^{S,obs} = \sum_{l=1}^{\infty} \sum_{m=1}^{l}
            \left[ |\stackrel{\rightarrow}{{\cal T}}_{ \omega_0 l}|^2
                  (\stackrel{\rightarrow}{W}^{S,em}_{lm}
                  -\stackrel{\leftarrow}{W}^{S,em}_{lm})
                  +\stackrel{\leftarrow}{W}^{S,em}_{lm}
            \right]
                  \;\; .
\label{WSOb}
\end{equation}
A numerical estimate of $W^{S,obs}/W^{S,em}$ is given by the solid line
in figure~\ref{fig05}.
In order to obtain an analytic approximation to the transmission coefficient
$|\stackrel{\rightarrow}{{\cal T}}_{ \omega_0 l}|^2$ and hence to
$W^{S,obs}$ (for $\omega^2_0/V_S^{\rm max} \ll 1$)
we first note that  asymptotically (\ref{Q}) reads
\begin{equation}
\stackrel{\rightarrow}{\psi^S_{ \omega_0  l}} (r)
\approx \frac{(2M)^{l+1} (l!)^2 }{(2l+1)! r^{l}}\ \ {\rm for}\ \ r \gg 2M\;\; .
\label{Qrlarge}
\end{equation}
Now, by fitting (\ref{Qrlarge}) with (\ref{->})
(with $\omega=\omega_0$)
in the range of $r$ satisfying $r\gg 2M$ and $\omega_0 r\ll 1$, we obtain
\begin{equation}
|\stackrel{\rightarrow}{{\cal T}}_{ \omega_0 l}| =
\frac{2^{2l+2} (l!)^3 (M \omega_0)^{l+1}}{(2l+1)! (2l)!} \;\; .
\label{T}
\end{equation}
The analytic approximation of $W^{S,obs}/W^{S,em}$ obtained in this way is
given by the dashed line in figure~\ref{fig05}.
It is seen that very little of the
emitted radiation is absorbed by the black hole.
This does not contradict the fact that Schwarzschild black holes
have a non-negligible absorption cross section
for infrared particles~\cite{P} (actually
of the order of the horizon area), because the main contribution
to the cross section comes from modes with $l=0$,
which are not emitted by our  circularly moving source.

\section{Final remarks}
\label{sec:final}
In summary, we have calculated the radiated power from a scalar source
rotating around a black hole in the framework of quantum field theory
at the tree level.
We have shown that for relativistic circular
orbits the emitted power is about 30\% and 20\% smaller
than what would be obtained in Newtonian gravity and in flat-spacetime
theory with gravitation generated by one-graviton exchange, respectively.
This is in agreement with the fact that
astrophysical processes  involving wavelengths of the order of the
Schwarzschild radius need to be described using fully curved spacetime.
We have also shown that most of the emitted energy escapes to  infinity.
Clearly, the presence of surrounding matter
could trap the emitted radiation in the vicinity of the black hole
because of friction effects~\cite{NY94}.
These astrophysical issues, however,  are beyond the scope
of the present paper.  Our conclusions are qualitatively in agreement
with those obtained for gravitational waves. The procedure used here 
can be readily adapted for scalar sources following other trajectories 
(see, e.g., \cite{L} for chaotic ones) by altering the scalar source 
(\ref{j}).

\acknowledgments
G.M. benefited from conversations with I. Novikov  and
W. Israel in the $9^{th}$ Brazilian School
on Gravitation and Cosmology. L.C. and G.M. would like to
acknowledge partial financial  support from CAPES
and CNPq, respectively, and S. Lietti for his computer
assistance. We thank E. Poisson for bringing one
of the papers in \cite{Pois} to our attention.

\section*{Appendix: A proof of Eq.\ (32)}

In this Appendix we prove the following formula:
$$
f(z) = \sum_{l=1}^\infty \sum_{m=1}^l m^2 [j_l(mz)]^2
|Y_{lm}(\pi/2,\phi)|^2 = \frac{1}{24\pi}\frac{z^2}{(1-z^2)^3}\ \ \ 
(|z| < 1)\;\;. 
\eqno{(32)}
$$
We note here that $|Y_{lm}(\pi/2,\phi)|^2 = |Y_{lm}(\pi/2,0)|^2$ is 
independent of $\phi$.

We start by showing that
\begin{equation}
f(z) = \frac{1}{32\pi^2}\int_0^{2\pi}d\phi\int_0^{\pi}d\theta\,\sin\theta\,
[g(z\sin\theta,\phi)]^2\;\;, \label{fzgz}
\end{equation}
where
\begin{equation}
g(a,\phi) \equiv \int_{-\infty}^{+\infty} d\psi\,
\delta'(\phi - \psi + a\sin\psi)\;\;, \label{defg}
\end{equation}
if $-1 < a < 1$. 
By combining the formulae
\begin{equation}
e^{ikz\cos\gamma} = \sum_{l=0}^\infty (2l+1)i^lj_l(kz){\rm P}_l(\cos\gamma)
\end{equation}
and
\begin{equation}
{\rm P}_l(\cos\gamma) = \frac{4\pi}{2l+1}\sum_{m=-l}^l
Y_{lm}(\theta',\psi)^* Y_{lm}(\theta,\phi)\;\;,
\end{equation}
where 
$\cos\gamma=\cos\theta\cos\theta'+\sin\theta\sin\theta'\cos(\phi-\psi)$,
and letting $\theta'= \pi/2$, we have
\begin{equation}
e^{ikz\sin\theta\cos(\phi-\psi)}
= 4\pi\sum_{l=0}^\infty \sum_{m=-l}^l
i^lj_l(kz)Y_{lm}(\pi/2,\psi)^*Y_{lm}(\theta,\phi)\;\;,
\end{equation}
By using the formula
$(2\pi)^{-1}\int_{-\infty}^{+\infty} d\psi\,e^{i\mu\psi} = \delta(\mu)$,
we find 
\begin{eqnarray}
& & \frac{1}{2\pi}\int_{-\infty}^{+\infty}d\psi\,
e^{ikz\sin\theta\cos(\phi-\psi)+ik\psi} \nonumber \\
& &  =   4\pi\sum_{l=0}^\infty \sum_{m=-l}^l i^l\, j_l(kz)
Y_{lm}(\pi/2,0)Y_{lm}(\theta,\phi)\delta(k-m)\;\;.
\end{eqnarray}
By multiplying by $k$ and integrating over $k$ we have 
\begin{eqnarray}
G(z,\theta,\phi) & \equiv & 
\frac{1}{8\pi^2}\int_{-\infty}^{+\infty} dk\int_{-\infty}^{+\infty}d\psi\,
k\,e^{ikz\sin\theta\cos(\phi-\psi)+ik\psi} \nonumber \\
& = &  \sum_{l=1}^\infty \sum_{m=1}^l i^l\,m\, j_l(mz)
Y_{lm}(\pi/2,0)[Y_{lm}(\theta,\phi)- Y_{l,-m}(\theta,\phi)]\;\;.
\label{Gztheta}
\end{eqnarray}
(We are using the convention 
$Y_{l,-m}(\theta,0) = Y_{lm}(\theta,0)$ here.)
Notice that this function is periodic in $\phi$ with priod $2\pi$.
By using orthonormality of the spherical harmonics $Y_{lm}(\theta,\phi)$
we find
\begin{equation}
f(z) = \frac{1}{2}
\int_{0}^{2\pi}d\phi \int_0^\pi d\theta\, \sin\theta\,
|G(z,\theta,\phi)|^2\;\;.  \label{Gzz}
\end{equation}
By shifting the integration variable in (\ref{Gztheta})
as $\psi \to \psi+\phi$, we have
\begin{eqnarray}
G(z,\theta,\phi) & = & 
\frac{1}{8\pi^2}\int_{-\infty}^{+\infty} dk \int_{-\infty}^{+\infty} d\psi\,
k\,e^{ik( z\sin\theta \cos \psi + \psi + \phi )}
\nonumber \\
& = & -i \frac{\partial\ }{\partial \phi}\left[ 
\frac{1}{8\pi^2}\int_{-\infty}^{+\infty} d\psi
\int_{-\infty}^{+\infty} dk e^{ik( z\sin\theta\cos\psi + \psi + \phi)}
\right]
\nonumber \\
& = & -\frac{i}{4\pi}\int_{-\infty}^{+\infty} d\psi\,
\delta'(z\sin\theta\cos\psi + \psi + \phi)\;\;. 
\end{eqnarray}
By changing the variable as $\psi \to \pi/2 - \psi$ we find 
\begin{eqnarray}
G(z,\theta,\phi) &= & -\frac{i}{4\pi}\int_{-\infty}^{+\infty}
d\psi\, \delta'(\phi + \pi/2 - \psi + z\sin\theta \sin\psi) \nonumber \\
& = & - \frac{i}{4\pi}\,g(z\sin\theta, \phi + \pi/2)\;\;,
\end{eqnarray}
where the function $g(a,\phi)$ is defined by (\ref{defg}). 
By substituting this in (\ref{Gzz}) and using the fact that $g(a,\phi)$ is
periodic in $\phi$ with period $2\pi$, we obtain (\ref{fzgz}).

Now, equation (\ref{fzgz}) can be rewritten as 
\begin{eqnarray}
f(z) & = & \frac{1}{32\pi^2}\int_0^{2\pi}d\phi\int_0^{\pi} d\theta\,\sin\theta\,
[g(z\sin\theta,\phi)]^2 \nonumber \\
& = & \frac{1}{32\pi^2}\int_0^{2\pi}d\phi\int_0^\pi d\theta\,\sin\theta\,
\int_{-\infty}^{+\infty} d\psi_1\,
\delta'(\phi-\psi_1 + z\sin\theta \sin\psi_1) \nonumber \\
& & \ \ \ \ \ \ \ \ \ \ \ \ \ \ \ \ \ \ \times 
\int_{-\infty}^{+\infty} d\psi_2\,
\delta'(\phi-\psi_2 + z\sin\theta \sin\psi_2) \nonumber \\
  & =  &  -\frac{1}{32\pi^2}
\int_0^{2\pi}d\phi
\int_0^{\pi}d\theta\,\sin\theta\, 
\int_{-\infty}^{+\infty} d\psi_1\,
\delta(\phi-\psi_1 + z\sin\theta \sin\psi_1) \nonumber \\
&&\ \ \ \ \ \ \ \ \ \ \ \ \ \ \ \ \ \times \int_{-\infty}^{+\infty} d\psi_2\,
\delta''(\phi-\psi_2 + z\sin\theta \sin\psi_2)\;\;. 
\end{eqnarray}
The function $\psi - z\sin\theta\sin \psi$ is a
monotonously increasing function of $\psi$ for $|z| < 1$, and the equation
$\phi = \psi - z\sin\theta\sin\psi$
is satisfied for $\phi = 0$ by $\psi = 0$ and for $\phi=2\pi$ by
$\psi = 2\pi$.  Hence, this equation for $\psi$ has a solution with 
a value of $\phi$ in $[0,2\pi]$ only if $0\leq \psi \leq 2\pi$.  Hence,
after performing the $\phi$ integral, the integration range for $\psi_1$
becomes $[0,2\pi]$.  Thus, we obtain
\begin{equation}
f(z) = - \frac{1}{32\pi^2}\int_0^\pi d\theta\,\sin\theta
\int_0^{2\pi}d\psi_1\int_{-\infty}^{+\infty} d\psi_2\,
\delta''(\chi_1- \chi_2)\;\;, 
\end{equation}
where $\chi_i = \psi_i - z\sin\theta\sin\psi_i$, ($i=1,2$).  Hence,
\begin{eqnarray}
f(z) &  = & -\frac{1}{32\pi^2}\int_0^\pi d\theta\,\sin\theta
\int_0^{2\pi}d\chi_1\,\frac{d\psi_1}{d\chi_1}\,
\int_{-\infty}^{+\infty}d\chi_2\,\frac{d\psi_2}{d\chi_2}
\,\delta''(\chi_1-\chi_2) \nonumber \\
& = & \frac{1}{32\pi^2} \int_0^\pi d\theta\,\sin\theta\,\int_0^{2\pi}d\psi\,
\left[ \frac{d^2\psi}{d\chi^2}\right]^2 \frac{d\chi}{d\psi} \nonumber \\
& = & \frac{z^2}{32\pi^2}\int_{0}^\pi d\theta\sin^3\theta
\int_0^{2\pi}d\psi\,\frac{\sin^2\psi}{(1-z\sin\theta\cos\psi)^5}\;\;,
\end{eqnarray}
where we have defined $\chi = \psi - z\sin\theta \sin \psi$.  
By integration by
parts with respect to $\psi$, we obtain
\begin{equation}
f(z) = \frac{z}{128\pi^2}\,h(z)\;\;, \label{final}
\end{equation}
where
\begin{equation}
h(z) = \int_0^\pi d\theta\,\sin^2\theta\int_0^{2\pi}
d\psi \frac{\cos\psi}{(1-z\sin\theta\cos\psi)^4}\;\;. \label{hz1}
\end{equation}
Now we evaluate this integral. 
Define
\begin{equation}
H(z) \equiv \frac{1}{3}\int_0^\pi d\theta\,\sin\theta
\int_0^{2\pi}\frac{d\psi}{(1-z\sin\theta\cos\psi)^3}\;\;. \label{defHz}
\end{equation}
The $\psi$ integral can be performed by using the residue theorem after
the substitution $t = e^{i\psi}$.  The result is
\begin{equation}
H(z)  =  \frac{2\pi}{3}\int_0^\pi\,d\theta\,\sin\theta
\left[ \frac{1}{(1-z^2\sin^2\theta)^{3/2}}
+ \frac{3}{2}\frac{z^2\sin^2\theta}{(1-z^2\sin^2\theta)^{5/2}}\right]\;\;.
\label{second}
\end{equation}
Then, by using the formulae
\begin{eqnarray}
\int \frac{dx}{(a+cx^2)^{3/2}} & = & \frac{1}{a}
\frac{x}{\sqrt{a+cx^2}}\;\;, 
\\
\int \frac{dx}{(a+cx^2)^{5/2}} & = & \frac{1}{a^2}
\left[ \frac{x}{\sqrt{a+cx^2}}
- \frac{cx^3}{3(a+cx^2)^{3/2}}\right] 
\end{eqnarray}
with $a = 1-z^2$ and $c=z^2$ in (\ref{second}) after the substitution
$x = \cos\theta$,
we have 
\begin{equation}
H(z) = \frac{4\pi}{3}\frac{1}{(1-z^2)^2}\;\;. \label{Hz}
\end{equation}
Noting that $h(z) = H'(z)$, we find from this
\begin{equation}
h(z) = \frac{16\pi}{3}\frac{z}{(1-z^2)^3}\;\;.
\end{equation}
By substituting this result
in (\ref{final}) we obtain equation (\ref{summation}).


\newpage

\newpage

\newpage

\newpage


\begin{thebibliography}{99}
\bibitem{misprint}
In \cite{HMSPRD98}, equation (7.3) and the one just above equation (7.5)
should read $ \psi_{0 l} (x) \approx C_2 \, (2l)! \, x^{l+1}/( l!)^2$
and  $C_2= (2^{2l+1} (l!)^3 M^{l+1} \, \omega^l )/((2l)! \, (2l+1)!) $
respectively. This does not affect other results in \cite{HMSPRD98}.
\end{thebibliography}
\end{document}